\documentstyle[aas2pp4]{article}
\received{March 24, 1998}
\accepted{May 1, 1998}
\lefthead{Keohane, Gotthelf \& Petre}
\righthead{Proportionate Partition in Cas\,A}

\begin{document}

\title{Evidence for Proportionate Partition  Between the Magnetic Field and
       Hot Gas in Turbulent Cassiopeia A}
\author{Jonathan W. Keohane\altaffilmark{1}, E. V. Gotthelf\altaffilmark{2} and R. Petre\altaffilmark{3}}
\affil{The Laboratory for High Energy Astrophysics \\
       Code 662, Goddard Space Flight Center  \\
       Greenbelt, MD \ 20771}

\authoraddr{Jonathan W. Keohane \hfill \linebreak Code 662,
NASA\discretionary{-}{}{/}GSFC \\ Greenbelt, MD \ 20771 }
\authoremail{jonathan@lheamail.gsfc.nasa.gov}

\altaffiltext{1}{The University of Minnesota; \linebreak[3] E-mail: jonathan@lheamail.gsfc.nasa.gov}
\altaffiltext{2}{Universities Space Research Association;  E-mail: gotthelf@gsfc.nasa.gov}
\altaffiltext{3}{E-mail:  petre@lheavx.gsfc.nasa.gov}

\vspace {0.5 cm}
\centerline{Accepted by ApJ Letters 5/1/98} 

\pagebreak[3]

\begin{abstract}
We present a deep X-ray observation of the young Galactic supernova remnant
Cas\,A, acquired with the ROSAT High Resolution Imager\@.  This high
dynamic range (232 ks) image reveals low-surface-brightness X-ray
structure, which appears qualitatively similar to corresponding radio
features.  We consider the correlation between the X-ray and radio
morphologies and its physical implications.   After correcting for the
inhomogeneous absorption across the remnant, we performed a point by
point (4\arcsec\ resolution) surface brightness comparison between the
X-ray and radio images.  We find a strong (r = 0.75) log-log
correlation, implying an overall relationship of $\log(\Sigma_{_{\rm
X-ray}}) \propto (2.21\pm0.05) \times \log(\Sigma_{_{\rm radio}})$\@.  This
is consistent with proportionate partition (and possibly equipartition)
between the local magnetic field and the hot gas --- implying that
Cas\,A's plasma is fully turbulent and continuously amplifying the
magnetic field.
\end{abstract}

% {\em PACS Codes: } 95.30.Qd, 95.75.Mn, 95.85.Nv, 98.38.Mz


\pagebreak[3]

\setcounter{footnote}{0}
\renewcommand{\thefootnote}{\fnsymbol{footnote}}

%\section{To Do}
%\begin{itemize}
%\item 
%\end{itemize}

\section{Introduction}
A comparison of the X-ray and radio emission of young supernova
remnants (SNRs) provides a powerful tool for investigating the physical
relationships among the thermal plasma, cosmic ray electrons and the
magnetic field (e.g., \markcite{1982ApJ...257..145D}{Dickel et al.\ 1982};
\markcite{1984ApJ...287..295M}{Matsui et al.\ 1984}; 
\markcite{1990ApJ...350..266A}{Arendt et al. 1990}; \markcite{1998ApJ...Dyer}
{Dyer \& Reynolds 1998})\@.  Radio emission is governed
primarily by the density of relativistic electrons and the magnetic
field strength, while the intensity in soft X-rays is dictated by the
gas density.   Shell SNRs provide excellent laboratories for
investigating the interaction among these physical processes, with
Cas\,A being the natural launching point for such investigations,
because of its high surface brightness in both the radio and X-ray
wavelength regimes.

Previous work has shown that the soft X-ray morphology of Cas\,A, on
angular scales $\gtrsim$30\arcsec\ (or 0.5 pc at a distance of 3.4
kpc), is dominated by absorption effects
(\markcite{1996ApJ...466..309K}{Keohane, Rudnick, \& Anderson 1996},
hereafter \markcite{1996ApJ...466..309K}{KRA})\@.  When this absorption
is taken into account, a higher degree of intrinsic correlation between
the X-ray and radio images is found.  Similarly, \markcite{1994PASJ...46L.151H}{Holt et al.\ (1994)} found a correlation
between the radio and the hard X-ray ($E$$>$$4.5$\,keV) surface
brightness, which is not affected by absorption.  These correlations
demonstrate that a simple relationship may exist among the underlying
physical parameters. 

In this Letter we investigate the relationship between the radio and
X-ray emission on smaller angular scales.  We use a newly acquired
ROSAT High Resolution Imager (HRI) image of Cas\,A, with a 40 times
longer exposure than used in \markcite{1996ApJ...466..309K}{KRA}\@. To
account for the inhomogeneous column density towards Cas\,A, we
``deabsorb'' the image using H{\sc~i} and OH absorption data
(\S\ref{deabsorption.sec})\@.  We perform a point by point
(4\arcsec\ resolution) surface brightness comparison between the X-ray
and radio images and find a statistically significant correlation
(\S\ref{radio_comparison.sec})\@.  We discuss the physical implications
of this correlation (\S\ref{discussion}) and its implications for
future observational and theoretical studies of Cas\,A and other young
SNRs (\S\ref{conclusion})\@.

%------------------------------------------------------------------------
%   Analysis Section

\pagebreak[3]
\section{Analysis}
\label{analysis}
%------------------------------------------------------------------------

\subsection{The Deep ROSAT Image of Cas\,A}
\label{raw_image.sec}
\begin{figure}
\epsscale{1.00}
\plotone {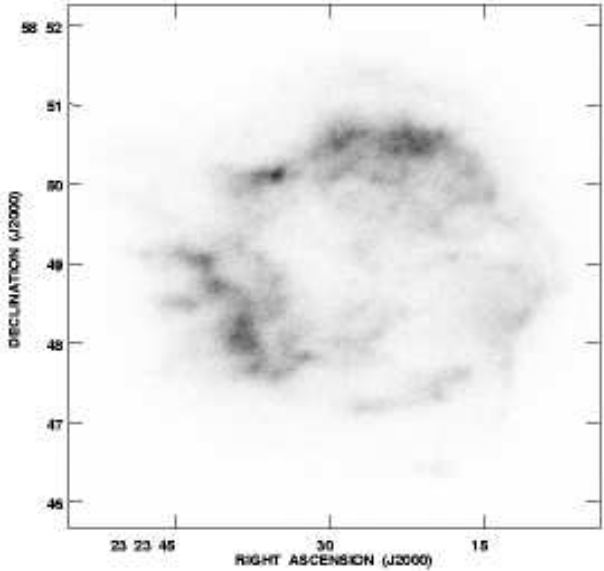}
\caption{The 232 ks ROSAT HRI raw image scaled linearly.  The maximum
value is $102 \rm \frac{counts}{0.5 \arcsec pixel}$\@.   This images is
oversampled with a pixel size of 0.5\arcsec.}
\label{raw_image}
\end{figure}

%------------------------------------------------------------------------
Deep ROSAT HRI exposures of the SNR Cas\,A were obtained on December 23,
1995 and June 20, 1996, with the telescope optical axis pointed toward
the center of the remnant.  The standard processing yields a total of
232\,ks of acceptable observing time, which produces the highest
dynamic range (500:1) X-ray image of Cas\,A to date. A final image,
shown in Fig. \ref{raw_image}, was made by cross-correlating and adding
images from the two exposures, using the radio image as an absolute
reference. 

In order to reduce instrumental blurring (e.g.\ the HRI ``halo'') we
applied 5 iterations of the Lucy-Richardson restoration technique
(\markcite{1972OSAJ...62...55R}{Richardson 1972};
\markcite{1974AJ.....79..745L}{Lucy 1974})\@.  As a kernel for the
deconvolution, we used an on-axis analytical radially symmetric point
spread function (PSF), which is valid because the PSF is approximately
constant across the 5\arcmin\ extent of Cas\,A\@.  The resolution of
the final image was estimated to be about 4\arcsec\ (0.07 pc).

%%%%  BEGIN FIGURES 

\begin{figure}
\epsscale{1.00}
\plotone {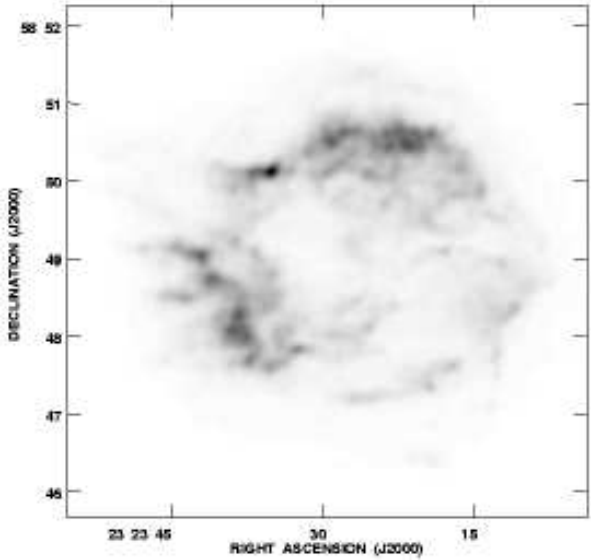} \\
\caption{The Lucy-Richardson deconvolved image with
4\arcsec\ resolution.  This greyscale is scaled linearly with a maximum
value of $5300 \rm \frac{counts}{4 \arcsec beam}$\@.   This image is
oversampled with a 0.5\arcsec\ pixel size, but has a resolution of
4\arcsec.}
\label{deconvolved_image}
\end{figure}

\begin{figure}
\epsscale{1.00}
\plotone {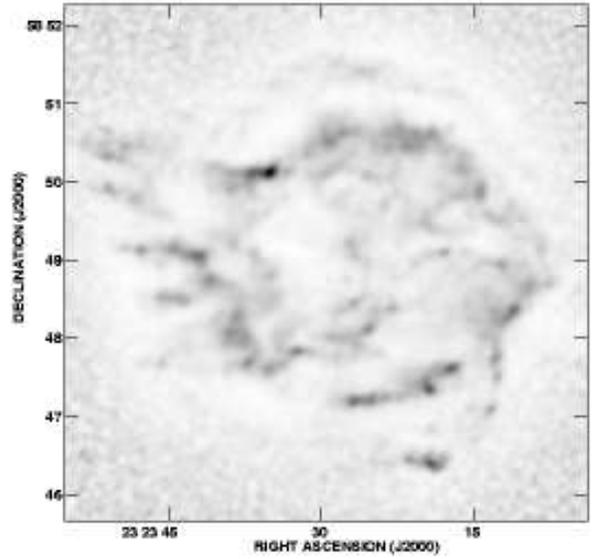} \\
\caption{The deconvolved ROSAT HRI image divided by a
60\arcsec\ smoothed version of itself.  This figure is included in
order to show the outlying and small scale structure revealed by this
deep exposure.}
\label{filtered_image}
\end{figure}

\begin{figure}
\epsscale{1.00}
\plotone {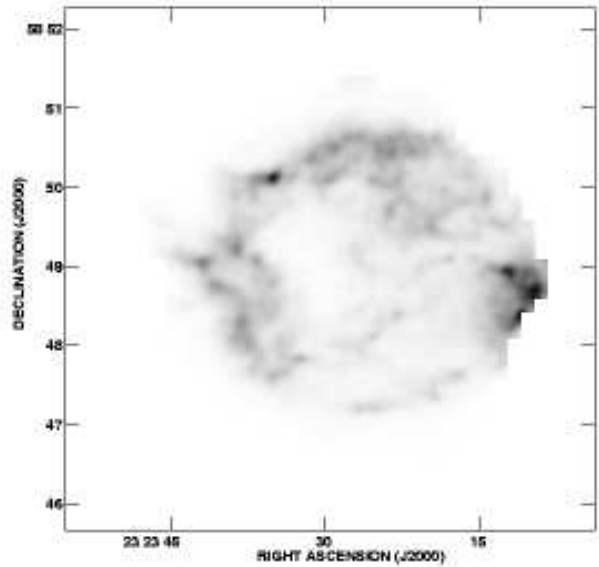} \\
\caption{The deabsorbed ROSAT image scaled linearly.   This image is
oversampled with a 0.5\arcsec\ pixel size, but its resolution is
4\arcsec\@.  Note that the intervening column density was mapped in
absorption, so this image is blanked outside of Cas\,A's radio-bright
disk where the H{\sc\,i} and OH optical depths are unreliable.}
\label{deabsorbed_image}
\end{figure}

%%%%  END FIGURES 

To correct any systematic scaling errors that may have been
produced by the restoration process, we used a 4\arcsec~FWHM Gaussian
smoothed version of the raw image to rescale the reconstructed image.
This required a linear transformation with a scale of 0.92 and an
insignificant zero-offset (0.31 cts/beam)\@.

This new image of Cas\,A is the first with enough sensitivity to reveal
X-ray features, on small angular scales ($\theta \gtrsim 4$\arcsec) and
low surface brightness ($\Sigma_{_{\rm HRI}} \gtrsim
0.03$\,cts\,s$^{-1}$\,(\arcmin)$^{-2}$), which have been only previously
detected in the radio.  Many of the most striking of these features are
outside the bright shell, such as the northern and southern knots and
the eastern jet region.  To enhance this outlying structure, we show a
spatially filtered version of the image (Fig.~\ref{filtered_image}),
produced by dividing the 4\arcsec\ X-ray image by a 60\arcsec\ Gaussian
smoothed version of itself.  This effectively removes large scale
structure and enhances regions away from the bright ring.

%------------------------------------------------------------------------

\subsection{The Deabsorbed Image}
\label{deabsorption.sec}
The large scale differences between Cas\,A's soft X-ray and radio
morphology are dominated by foreground absorption effects
(\markcite{1996ApJ...466..309K}{KRA})\@.   In order to take these
effects into account, we used H{\sc~i} and OH absorption data
(\markcite{1997A&AS..123...43S}{Schwarz, Goss, \& Kalberla 1997};
\markcite{1986ApJ...310..853B}{Bieging \& Crutcher 1986}), which trace
the foreground atomic and molecular gas respectively.  We translated
these data into an optical depth map using the empirical relation
(specific to Cas\,A) for each resolution element:
\begin{equation}
\label{tau.eqn} \tau_{_{X}} = 0.073 \, {\rm  km^{-1}s } \times EW_{\rm
H{\sc\,i}} +
	      1.6 \, {\rm km^{-1}s} \times EW_{\rm OH} + 1.6
\end{equation} 
where $\tau_{_{X}}$ is Cas\,A's optical depth over the ROSAT HRI
bandpass, and $EW_{\rm H{\sc\,i}}$ and $EW_{\rm OH}$ are the H{\sc~i}
and OH equivalent widths respectively.  A complete description of this
method is given in \markcite{1996ApJ...466..309K}{KRA}\@.

Using this HRI-specific 30\arcsec\ resolution optical depth map, we
produced the deabsorbed
X-ray image of Cas\,A shown in Fig.~\ref{deabsorbed_image}\@.   Absorption
structure on scales between 30\arcsec\ and 4\arcsec\ could affect our
results, but the OH data of 
\markcite{1986ApJ...310..853B}{Bieging \& Crutcher} suggest that
clouds of scales 30\arcsec--60\arcsec\ dominate the absorption
structure.

%------------------------------------------------------------------------

\subsection{A Comparison with the Radio Image}
\label{radio_comparison.sec}

% figs 4 & 5 go back here for preprint

\begin{figure*}[t]
\epsscale{1.75}
\plotone{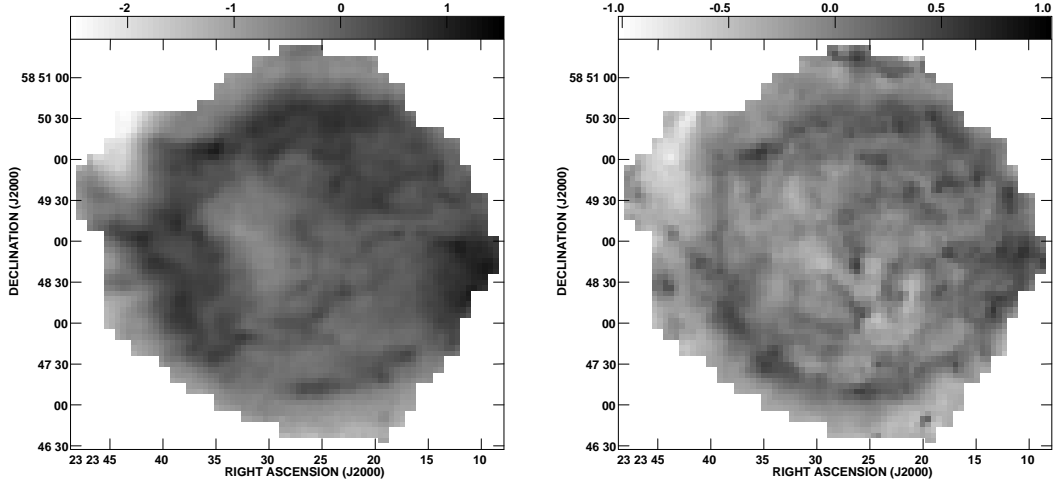} \\
\epsscale{1.75}
\caption[Logarithmically scaled images of Cas\,A]
{Logarithmically scaled images of Cas\,A\@.  
We present, from left to right, the deabsorbed X-ray image
and the $\lambda6$~cm VLA map.  The pixel size is 4\arcsec,
the same as the beam size.  The images are scales to correspond
to the axes in Fig~\ref{radio_v_Xray}\@.}
\label{log_images}
\end{figure*}

\begin{figure}[tbp]
\epsscale{0.90}
\plotone {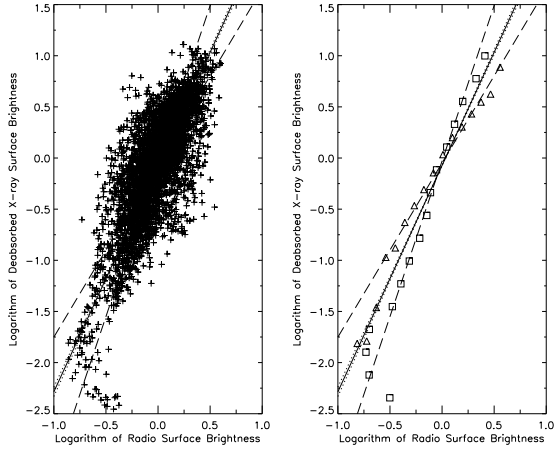} \\ 
\epsscale{1.00}
\caption[X-ray vs.\ radio scatterplot of Cas\,A]{Log-log plots
correlating the images in Fig~\ref{log_images}\@.  The left-hand plot
is a standard scatter-plot with 3948 data points.  For a majority of
points, the random errors are smaller than the symbols plotted.  On the
right is a plot of the median $\log(\Sigma_{_{\rm X\,ray_{D}}})$ in
evenly spaced bins of $\log(\Sigma_{_{\rm radio}})$ (triangles) and
visa versa (squares)\@.  Both data sets are scaled to units of their
respective overall median values.  The same lines are plotted on each
graph:  the dashed lines represent the two standard least-squares fits
($\eta_{\rm xy}$=1.7, $\eta_{\rm yx}$=3.0);
the dotted lines represent the 90\% confidence range in the slope; and
the solid line is the bisector of the two least-squares slopes. }
\label{radio_v_Xray}
\end{figure}

The brightest features in the X-ray and radio maps are for the most
part not correlated.  However, when viewed on logarithmic scales, the
deabsorbed X-ray map and the radio map appear quite similar
(Fig.~\ref{log_images})\@.  In order to quantify this similarity, we
performed a point-by-point comparison between the deabsorbed X-ray
image and the most recent (epoch 1994) $\lambda6 \, \rm cm$ VLA map
(\markcite{1998..............K}{Koralesky \& Rudnick 1998})\@.  The
time differences between the radio and X-ray observations
($\sim$1~year) will have no effect, given that typical proper motions
in Cas\,A are of order $0.1 \arcsec \rm \, yr^{-1}$
(\markcite{1986MNRAS.219...13T}{Tuffs 1986})\@.
Figure~\ref{radio_v_Xray} shows this correlation for 4\arcsec\ pixels.
This relationship has a correlation coefficient of r$=$0.75, and
appears to be real but scatter-dominated.

We can parameterize the relationship between the deabsorbed X-ray
surface brightness ($\Sigma _{\rm X\,ray_{D}}$) and the radio surface
brightness ($\Sigma _{\rm radio}$) as
\begin{equation}
\label{eta_def.eqn}
\Sigma _{\rm X\,ray_{D}} \propto \Sigma _{\rm radio}^{\eta} ~~.
\end{equation}
To find the value of $\eta$, we performed a least-squares analysis
between the quantities $\log(\Sigma _{\rm X\,ray_{D}})$ and
$\log(\Sigma _{\rm radio})$\@.  We found the bisector between the two
standard least-squares fits, as is appropriate for a scatter dominated
linear correlation between two physical quantities
(\markcite{1990ApJ...364..104I}{Isobe et al.\ 1990)}\@.  At 4\arcsec\ binning we find a
best-fit $\eta=2.21$ with a corresponding variance of $\sigma_\eta^2 =
6.8 \times 10^{-4} = (0.026)^2$\@.   Monte Carlo simulations by 
\markcite{1990ApJ...364..104I}{Isobe et al.}\ suggest that their
formula for finding the variance in the slope underestimates the true
uncertainty by about 10\%, so we have set our ``90\% confidence''
errors (Table~\ref{eta_scale.tab}) to the $1.8\sigma$ level\@.

We performed two simple tests to show that neither the correlation nor
the scatter are artifacts of our analysis.  First, to show that the
correlation is intrinsic to the data, we rotated one image by
180\arcdeg\ about Cas\,A's geometrical center before re-correlating the
data, whereby the correlation coefficient dropped from $r$=0.75 to
$r$=0.41\@.  This implies that the correlation is due to small scale
structure, and is not simply a byproduct of Cas\,A's overall circular
morphology.  Second, to show the reality of the scatter, we created a
simulated X-ray map, by assigning to each pixel an X-ray surface
brightness that combined the radio surface brightness of that pixel
scaled by a power law of 2.21, plus a random term representing the
Poisson scatter in the X-ray map.  The correlation between the radio
and the simulated X-ray image showed no significant scatter, implying
that the scatter in figure~\ref{radio_v_Xray} due to statistical
fluctuations is small compared with that observed.  Possible physical
explanations for this scatter are discussed in \S\ref{scatter_dis}\@.

We performed the correlation analysis described above on maps with
successively larger bin sizes in order to investigate the robustness of
our analysis and investigate how increasing the pixel size affects the
value of $\eta$ (Table~\ref{eta_scale.tab})\@.  As the data are binned,
the best-fit $\eta$ decreases, but its maximum value remains at about
$\eta_{\rm max} \approx 2.3$\@. 

To illustrate how smoothing could bias the measured values of $\eta$,
assume that there exists a value $\eta_{\circ}$$>$$1$, such that $\Sigma
_{\rm X\,ray_{D}} = C \times \Sigma _{\rm radio}^{\eta_{\circ}}$\@.  Thus
the averaged quantities would be subject to the inequality:
\begin{equation}
\left \langle \Sigma _{\rm X\,ray_{D}} \right \rangle  =   \left
\langle C \times  \Sigma _{\rm radio}^{\eta_{\circ}}  \right \rangle  \ge  C
\times  \left \langle \Sigma _{\rm radio} \right \rangle^{\eta_{\circ}} \, .
\label{eta_inequality.eqn}
\end{equation}
One possible explanation for the decreasing $\eta$ with resolution is
that this bias affects both the dim and bright parts of Cas\,A to about
the same degree; however in proportion to their average brightness the
dimmer regions are affected to a greater degree, so our logarithmic
fitting returns a flatter slope $\eta$\@.   Another possibility is that
the binning of data lowers the dynamic range of the fit, thus skewing
$\eta$ towards its uncorrelated value of unity.  A third possible
explanation is that smoothing increases the importance of path-length
variations as the governing physical parameter, thus also skewing
$\eta$ towards unity.  Regardless of the particular explanation, the
trend shown in Table~\ref{eta_scale.tab} suggests that the intrinsic
value of $\eta$ is between 2.2 and 2.3\@.

%%%%  BEGIN FIGURES 

\begin{figure}[tbp]
\epsscale{1.00}
\plotone {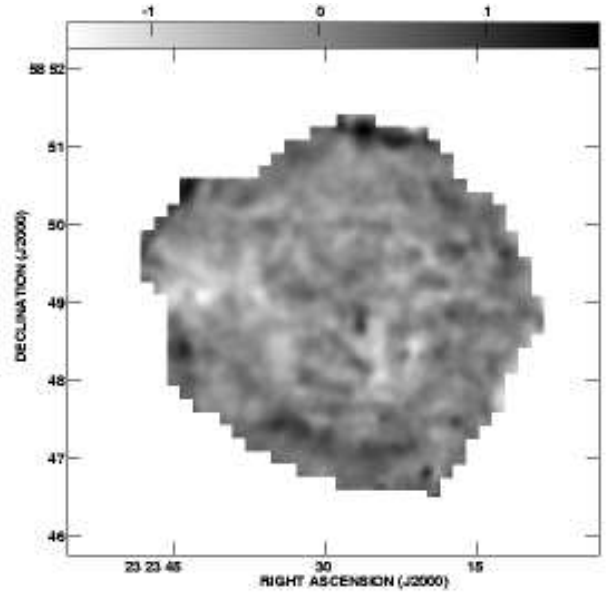} \\ 
\caption[Image of the residuals from Cas\,A's X-ray/radio correlation]
{Plot of the quantity $\log(\Sigma_{_{\rm
radio}}^{2.21}/\Sigma_{_{\rm X\,ray_{D}}})$\@.  The greyscale ranges from
$\log(\rm min/mean) = -1.51$ (white) to $\log(\rm max/mean) = 1.67$
(black)\@.  Like Fig.\ref{deabsorbed_image}, this image is oversampled
with a 0.5\arcsec\ pixel size, but a resolution of 4\arcsec.}
\label{residuals}
\end{figure}

%%%%  END FIGURES 

%------------------------------------------------------------------------

\begin{table}[t]
\caption{The X-ray/radio Regression Parameters}
\begin{tabular}{||r|r|c|c|||}\hline\hline
Resolution & N~~ & r & $\eta$  \\ \hline
4\arcsec  & 3948& 0.75 & $2.21 \pm 0.05$  \\
8\arcsec  & 876 & 0.75 & $2.19 \pm 0.11$ \\
16\arcsec & 210 & 0.76 & $2.14 \pm 0.17$ \\
32\arcsec & 47  & 0.75 & $1.96 \pm 0.35$  \\
64\arcsec & 10  & 0.49 & $1.43 \pm 0.72$ \\ \hline\hline
\end{tabular} \\
N = the number of independent points  \\
r = the correlation coefficient \\
$\eta$ = the log-log surface brightness slope (see equation~\ref{eta_def.eqn}) with 90\% confidence errors \\
\label{eta_scale.tab}
\end{table}

\subsection{Deviations from the Trend}
\label{analysis_deviations.sec}
Figure~\ref{residuals} is a map of the quantity $\log(\Sigma_{_{\rm
radio}}^{2.21}/\Sigma_{_{\rm X\,ray_{D}}})$, or a measure of deviations
from the best-fit slope $\eta$\@.  Consistent with an overall physical
correlation, this map shows very little structure.  There are no large
scale features such as the shell or the western hot spot.  A few key
features are present, however.  Some are relatively radio bright:  the
northern knot region; a knot near the center; and the outlying
structure in the south.  In addition, there is a relatively X-ray
bright feature on the eastern shell.

%  Figure residuals goes back here for preprint

%------------------------------------------------------------------------
%   Discussion Section
\pagebreak[3]

\section{Discussion}
\label{discussion}  

\subsection{Proportionate Partition as a Possible Origin for the Correlation}

   The main objective of comparing the X-ray and radio emission in
Cas\,A is to identify the physical relationships among the thermal
plasma, cosmic ray electrons and magnetic field.  In this section we
show that our results are consistent with proportionate partition
between the energy content of the gas and magnetic field.  We also
develop a physical justification for this by postulating a turbulent
magnetohydrodynamic (MHD) plasma, which we argue is continuously
amplifying the magnetic field.  As both the trend and the exceptions
are real, this scenario should be interpreted as valid ``on average,''
but not necessarily for each feature in Cas\,A\@.

The ROSAT band spectrum of Cas\,A is dominated by
line emission from H and He-like ions of Mg and Si and the L-shell ions
of Fe (\markcite{1994PASJ...46L.151H}{Holt et al.\ 1994})\@.  Thus by
assuming a constant temperature plasma, we can relate the X-ray
emissivity\footnote{Here the emissivity, $\varepsilon_{_{\rm X\,ray}}$,
is defined in an analogous way to the radio emissivity, such that
$\Sigma_{_{\rm X\,ray_{D}}} \propto \int  \varepsilon_{_{\rm X\,ray}}
dl$, where $l$ is the path length.} ($\varepsilon_{_{\rm X\,ray}}$) to
the gas density ($n_{\rm gas}$) as:
\begin{equation}
  \label{xray_emissivity.eqn}
   \varepsilon_{_{\rm X\,ray}} \propto n_{\rm gas}^2
\end{equation}

Cas\,A's radio emission mechanism is believed to be synchrotron radiation
(\markcite{1970MNRAS.151..109R}{Rosenberg 1970}), arising from relativistic
electrons (of density $n_{\rm rel}$) interacting with the magnetic field
($B$)\@.  We therefore characterize the radio emissivity
($\varepsilon_{_{\rm radio}}$) as:
\begin{equation}
  \label{radio_emissivity.eqn}
  \varepsilon_{_{\rm radio}} \propto n_{\rm rel} B^{1+\alpha}
\end{equation}
where $\alpha$ is the radio spectral index ($\alpha =
0.77$, \markcite{1977A&A....61...99B}{Baars et al.\ 1977})\@.

\markcite{1994ApJ...421L..31A}{Anderson et al.\ (1994)}
concluded that the radio brightness of the compact features in Cas\,A is governed primarily by
magnetic field amplification.  For this reason, we believe it
reasonable to assume a constant (or uncorrelated) relativistic
electron density.  Therefore, equation
\ref{radio_emissivity.eqn} becomes:
\begin{equation}
  \label{radio_emissivity_simple.eqn}
  \varepsilon_{_{\rm radio}}  \propto B^{1+\alpha}
\end{equation}   

In their MHD simulations, \markcite{1995ApJ...453..332J}{Jun, Norman,
\& Stone (1995)} show that Rayleigh-Taylor (R-T) instabilities in young
SNRs are significantly enhanced by the presence of a magnetic field.
Kelvin-Helmholtz instabilities, created by the shear flow along the R-T
fingers, greatly enhance the magnetic field --- especially on small
scales.  In addition, MHD studies of simple shear systems show that the
magnetic field is amplified quickly as the energy cascades to smaller
scales \markcite{1996ApJ...456..708M}{(Malagoli, Bodo, \& Rosner
1996)}\@.  On the scale where the magnetic field becomes dynamically
important, reconnection and dissipation occur.  It is therefore
plausible that the magnetic field in Cas\,A is being amplified to the
point of equipartition with the hot gas, locally on the smallest
observable scales, so:
 \begin{equation} \label{equipartition.eqn} \frac{B^2}{8\pi} \approx
U_{\rm gas} \end{equation} In aggregate this is approximately true.  If
one assumes equipartition between the magnetic field and the
relativistic plasma, an energy density for each of these can be derived
simply from the overall radio flux density.  This overall magnetic
energy density ($\frac{B^2}{8\pi}$) is comparable to the thermal energy
density ($U_{\rm gas}$) estimated from the integrated ASCA X-ray
spectrum (Allen, G.E. {\em private communication})\@.

This suggestion of equipartition is apparently incompatible with the
simulation of \markcite{1996ApJ...472..245J}{Jun \& Norman (1996)},
whose 3-D simulations predicted magnetic fields significantly lower
than equipartition.  However, they also found that the higher the
resolution of the simulation (i.e., the lower effective viscosity), the
greater the resultant magnetic field.  With regard to the analysis
presented here, this distinction is moot.  If the turbulent cascade is
halted on the smallest scales by the magnetic pressure, the
equipartition relation (equation~\ref{equipartition.eqn}) should hold.
On the other hand, if the turbulent cascade is the primary amplifier of
the magnetic field, but is halted before the magnetic field reaches
equipartition, then one would expect the system to still be in
proportionate partition, i.e.:
\begin{equation}
\label{proportionate_partition.eqn} B^2 \propto U_{\rm gas} \propto n_{\rm gas} 
\end{equation} 
The analysis presented here cannot distinguish between equipartition
(equation~\ref{equipartition.eqn}) and the looser constraint of
proportionate partition (equation~\ref{proportionate_partition.eqn})\@.  Similarly, it does not matter
here whether $U_{\rm gas}$ represents the turbulent energy density or
the thermal energy density, so long as the energy density scales
linearly, on average, with the gas density ($U_{\rm gas} \propto n_{\rm
gas}$)\@.  We now combine equations \ref{xray_emissivity.eqn},
\ref{radio_emissivity_simple.eqn} and \ref{proportionate_partition.eqn}
to find a relationship between the X-ray and radio emissivities:
\begin{equation}
\label{x-ray_radio_emissivity.eqn}
  \varepsilon_{_{\rm X\,ray}}  \propto  B^{4}  \propto  
         \varepsilon_{_{\rm radio}}^{\frac{4}{1+\alpha}} \propto
         \varepsilon_{_{\rm radio}}^{2.26}
\end{equation}
This simple relationship agrees remarkably with the value of
$\eta$ found in \S\ref{radio_comparison.sec}\@. 

A similar and independent conclusion about turbulent magnetic field
amplification was found by 
\markcite{1998..............K}{Koralesky \& Rudnick (1998)}, who
examined the changes in Cas\,A's radio emission as a function of
azimuth over an 11 year baseline.  They found that regions with the
largest decrease in fractional polarization (corresponding to a
decrease in order of the magnetic field) are decreasing the least in
flux density.  They suggest that the same process which is randomizing the
magnetic field's orientation is also amplifying it, partially
compensating for expansion losses.

This relationship between the radio and X-ray surface brightness is
very different from the predictions of most other simple models.  For
example, in an isobaric model, the magnetic pressure and thermal
pressure add to a constant, so the surface brightnesses would be
anti-correlated.  More realistically, a compression scenario where the
magnetic field scales inversely with area so $B \propto n_{\rm
gas}^{2/3}$, implies that $\eta = 1.7$, which is also inconsistent with
the data.  If instead, we make the equally simple assumption that the
relativistic electron density is proportional to the thermal gas
density ($n_{\rm rel} \propto n_{\rm gas}$), then we find $\eta =
2.0$\@.  While we cannot completely rule out this possibility, we do
not favor it.  The Larmor radii of the relativistic electrons are very
small ($\sim$$10^{9}$\,cm for a 1\,GeV particle in a 1\,mG magnetic
field), tightly coupling them to magnetic field lines, which in turn
are coupled to the plasma.  This suggests that $n_{\rm rel} \propto
n_{\rm gas}$ may be a good assumption, but then one would also expect a
relation between the magnetic field and density.  When a magnetic field
dependence is included, $\eta$ is significantly reduced.  For example,
if $n_{\rm rel} \propto n_{\rm gas} \propto B$ then $\eta = 0.7$, which
is clearly inconsistent with the correlation.
 
\subsection{Possible Contributions to the Scatter}
\label{scatter_dis}
The simple model that fits the surface brightness ratio rests on three
assumptions:  constant relativistic electron density, constant X-ray
emissivity per density squared ($\varepsilon_{_{\rm X\,ray}}/n^2_{\rm
gas}$), and proportionate partition.  The relative
flatness of the ratio map shows that deviations from these assumptions
on large scales are absent.  It is plausible, however, that the
substantial scatter of the ratio for individual cells from the trend
line represents local violation of one or more of these assumptions.
The log of the rms deviation about the trend line is $\pm$0.38 (a
factor of $\sim$2.5); the log of the maximum excursion is $\pm$1.5.

If the scatter is due entirely to variations of $n_{\rm rel}$ or
$B^2/n_{\rm gas}$ (proportionate partition), we find the corresponding rms
variation of these is $\pm$0.17 in the log, or a factor of $\sim$1.5.
The X-ray emissivity is a function of both gas temperature and metal
abundance.  The emissivity scales linearly with the abundance of the
metals producing the dominant lines in the ROSAT band, namely Mg and
Si; thus local rms variations of a factor of $\sim$2.5 in metallicity
can account for the scatter.  The emissivity scales with temperature in
a less straightforward way, but deviations of $\log(\varepsilon_{_{\rm
\tiny X\,ray}}/n^2_{\rm gas})$ of $\pm$0.4 can be accounted for by
log(kT) variations of $\pm$0.4.

Scatter of the degree indicated in any of these parameters is
physically plausible.  For instance, small scale gas pressure
variations associated with a shock encountering a density enhancement
can be as large as a factor of 6 
(\markcite{1996ApJ...471..279D}{Dohm-Palmer \& Jones 1996}), 
which in turn could produce variations of B$^2/n_{\rm gas}$ or
$\varepsilon_{_{\rm X\,ray}}/n^2_{\rm gas}$ within the observed range.
Measurements of any of these parameters on small angular scales, at best
pose a challenge to current and future instruments, and at worst may be
infeasible to measure (e.g., n$_{\rm rel}$)\@.

\section{Conclusion}
\label{conclusion}
We have performed a comparison of Cas\,A's radio and X-ray emission to
a limiting resolution of 4\arcsec\ (0.07 pc)\@.  The strong correlation
between radio and X-ray surface brightness can be explained by the
scenario that Cas\,A has ``on average'' a spatially constant
relativistic electron density and proportionate partition on small
scales between its thermal and magnetic energy densities, as would be
expected from a fully turbulent MHD system.
 
These results may have important implications for theoretical work.
The complex plasma physics of SNRs often must be simplified in order
for computer simulations to run in a cost-effective manner.
Unfortunately, most often one assumption is spherical or cylindrical
symmetry, which is incompatible with turbulent flow.   It may be
possible, instead, to use the proportionate partition relation
suggested by our result, along with some other characteristics of fully
turbulent systems, to produce a more realistic first-order description
of young, core-collapse, supernova remnants.

\acknowledgments
We thank Barron Koralesky for kindly providing the $\lambda6$~cm radio
data of Cas\,A in a digitized form.  We also wish to thank Glenn Allen
and John Dickel for insightful conversation.  This work was supported
by NASA.

%%%%  BEGIN REFS

%%%%  END  REFERENCES  

\end{document}